\documentclass[a4paper,10pt]{article}
\usepackage[a4paper,total={16cm,24.7cm}]{geometry}
\usepackage[utf8]{inputenc} 
\usepackage{indentfirst}
\usepackage{amsmath}
\usepackage{graphicx}
\usepackage{multirow}
\usepackage{subfigure}
\usepackage{float}
\usepackage{authblk}
\usepackage[small,bf]{caption}
\usepackage{cite}

\providecommand{\keywords}[1]
{
	\small	
	\textbf{\textit{Keywords:}} #1
}

\title{The inverse problem for fractal curves solved with the dynamical approach method}

\author[1,2a]{Luiz Bevilacqua\footnote{Corresponding author. E-mail: bevilacqua@coc.ufrj.br}} 
\author[2a,2b,2c]{Marcelo M. Barros} 
\author[2a,2b]{Felipe C. V. Venturelli} 

\affil[1]{\small{Coimbra Institute. Graduate School of Engineering, COPPE - Universidade Federal do Rio de Janeiro, Brazil}}
\affil[2a]{Laboratory for Interdisciplinary Numerical Modeling, LIMON -  Universidade Federal de Juiz de Fora, Brazil}
\affil[2b]{Graduate School of Civil Engineering, PEC - Universidade Federal de Juiz de Fora, Brazil} 
\affil[2c]{Department of Structural Engineering, ETU - Universidade Federal de Juiz de Fora, Brazil}

\date{} 

\begin{document}
	\maketitle
	
	\begin{abstract}
		\noindent 
		The purpose of the present paper is to present the main applications of a new method for the determination of the fractal structure of plane curves. It is focused on the inverse problem, that is, given a curve in the plane, find its fractal dimension. It is shown that the dynamical approach extends the characterization of a curve as a fractal object introducing the effects of mass density, elastic properties, and transverse geometry. The dynamical dimension characterizes material objects and suggests that biological characterization can be much more complete with the methodology presented here. 
	\end{abstract} \hspace{5pt}

	
	
\section{Introduction}
	
	The geometry of curves in a plane is usually analyzed with the help of mathematical measures theories. Particularly important in recent years is the theory introduced by Hasdorff aimed at exploring the fractal dimension of a certain class of curves, namely fractal curves \cite{Falconer}. Let us call fractal a sequence of curves, called fractal sequence, composed by curves $F_i$ with length $L_i$ such that $\lim_{i\rightarrow \infty} L_i \rightarrow \infty$, that can be packed into a two-dimensional box $B$ and it will always be possible to find a finite neighborhood $\Omega \in B$ such that $\lim_{i \rightarrow \infty} F_i \cap \Omega = \emptyset$. That is, a fractal curve no matter its length will never fill the box. The success of fractal curves can partly be explained by their practical applications in representing the geometry of objects found in nature \cite{Mandelbrot,Feder,Turcotte}. The association of biological functions with fractal dimensions has stimulated research efforts to improve knowledge in this area \cite{Bassingthwaighte}. Despite considerable advances, a specific issue needs to be improved namely the consideration of complementary variables in real objects such as mass, thickness, and physical properties. This paper introduces a method capable of taking into account material and geometric properties that cannot be achieved with the classical approach to fractal objects.
	
\begin{figure}[h]
		\centering
		\includegraphics[width=0.9\textwidth]{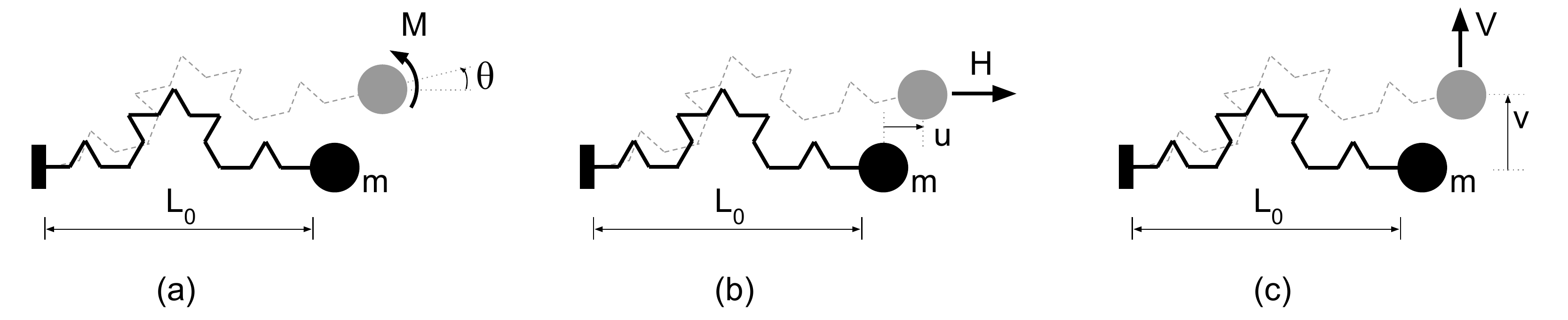}
		\caption{Simple oscillator corresponding to the second term of the Koch triadic. The elastic spring is bent due to: (a) the action of a moment, (b) the action of a horiontal force or (c) the action of a vertical force, as initial conditions.}
		\label{fig:IC}
	\end{figure}

	The method explored in this paper is derived from the dynamical behavior of simple mass-spring oscillators whose spring is an elastic wire bent with the same shape as the fractal curve. A typical harmonic oscillator is shown in Fig.\ref{fig:IC}. The wire is fixed at the left end and carries a mass $m$ at the right end. Three different displacements can be independently applied to the mass, as shown in the figure. Therefore, each spring-mass oscillator exhibits three independent natural frequencies. Each frequency can be excited by an initial displacement induced by a moment $M$, a vertical force $V$ or a horizontal force $H$. The frequency response of simple oscillators assembled as proposed here leads to a deeper source of information as compared with purely geometric approaches.
	
	Consider a curve obtained by assembling a collection of elements all with the same size, to obtain a single chain with no bifurcations. The general expression for the period corresponding to the $k^{\text{th}}$ term in the sequence for any of the three inicial conditions in Fig.\ref{fig:IC} can be shown to be \cite{Bevi0,Bevi1}:

	\begin{equation}
		\frac{T_k^2}{T_0^2} = N_k \frac{\lambda_k}{L_0} \Omega_k,
		\label{eq:Tk}
	\end{equation}
	where $T_0^2=m_0h_0^2L_0/E_0I_0$ is a normalization constant obtained with material and geometric characteristics of a reference oscillator. For the initial excitation induced by a moment $M$, the factor $\Omega_k=1$ and, the equation \eqref{eq:Tk} can be rewritten:
	
	$$
	\log \left(\frac{T_k}{T_0}\right) = \frac{1}{2} \left(\log N_k + \log \frac{\lambda_k}{L_0}\right),
	$$
	or:
	
	$$
	\log \left(\frac{T_k}{T_0}\right) = \frac{1}{2} (-F_M+1)\log\frac{\lambda_k}{L_0} \qquad \text{where} \qquad
	\log N_k = -F_M \log(\lambda_k/L_0).
	$$
	
	Now from the relation above we have $N_k(\lambda_k)^{F_M}=(L_0)^{F_M}$ and clearly $F_M$ is the fractal dimension of the curve. Therefore, the equation \eqref{eq:Tk} can be reduced to the following expression:
	
	\begin{subequations}
	
	\begin{equation}
		\log\left(\frac{T_k}{T_0}\right)=\frac{1-F_M}{2}\log\frac{\lambda_k}{L_0}
		\label{eq:logTkM}
	\end{equation}

	The equation \eqref{eq:logTkM} is a fundamental equation to determine the fractal dimension of plane curves with the dynamical response of simple oscillators.
	
	The other two independent initial displacements may be used to determine the complementary periods of the harmonic oscillators, namely the displacement induced by a horizontal force and the displacement induced by a vertical force, Fig.\ref{fig:IC}(b-c). The expressions are similar, except that the factor $\Omega_k \ne 1$, then the general expression for the relative period is:
	
	\begin{equation}
		\log\left(\frac{T_k}{T_0}\right) = \frac{1-F_{H,V}}{2} \log\frac{\lambda_k}{L_0}+\log \Omega_k
		\label{eq:logTkHV}
	\end{equation}

	The extra term $\Omega_k$ in \eqref{eq:logTkHV} comes from the position and orientation of the elements $\lambda_k$ in the geometry. Indeed, the bending moment distribution induced by the moment $M$, Fig. \ref{fig:IC}(a), does not depend on the position and orientation of the elements $\lambda_k$. Therefore, the bending moment is uniformly distributed along the spring length leading to $\Omega_k=1$. For the other cases, corresponding to the initial displacement induced by a horizontal or vertical force, the bending moment distribution and consequently the displacement of the mass $m$ depends on the position and orientation of the elements $\lambda_k$ in the spring geometry. This non-homogeneous distribution of the bending moment leads to the inclusion of a disturbance term $\Omega_k$ in the determination of the period of oscillation. It can be shown that for regular fractal curves the term $\Omega_k\rightarrow 1$ for large values of $k$ \cite{Bevi0,Bevi1,Bevi2,Bevi3}. Therefore the angular coefficients of the straight lines corresponding to the functions $\log(T_k/T_0)\times \log(\lambda_k/L_0)$ for the three initial conditions $M, H$ and $V$ are related with the Hausdorff dimension as $D_M=D_H=D_V=(1-F)/2 $ since all three angular coefficients coincide for the cases under consideration, $F_M=F_H=F_V=F$.
	
	Therefore, the fractal dynamical dimension depends on the elastic energy distribution in the oscillator spring. The period of oscillation excited by a moment $M$ is independent of the orientation of the elements $\lambda_k$. Now, the elastic energy stored in the system generated by the bending moment is a function of the position and orientation of each element in the curve. This property is particulary important for detecting random assemblages of fractal curves. The random assemblages as shown in Fig.\ref{fig:rand} can be characterized with the dynamical fractal method.
	
	\begin{figure}[ht]
		\centering
		\includegraphics[width=0.7\textwidth]{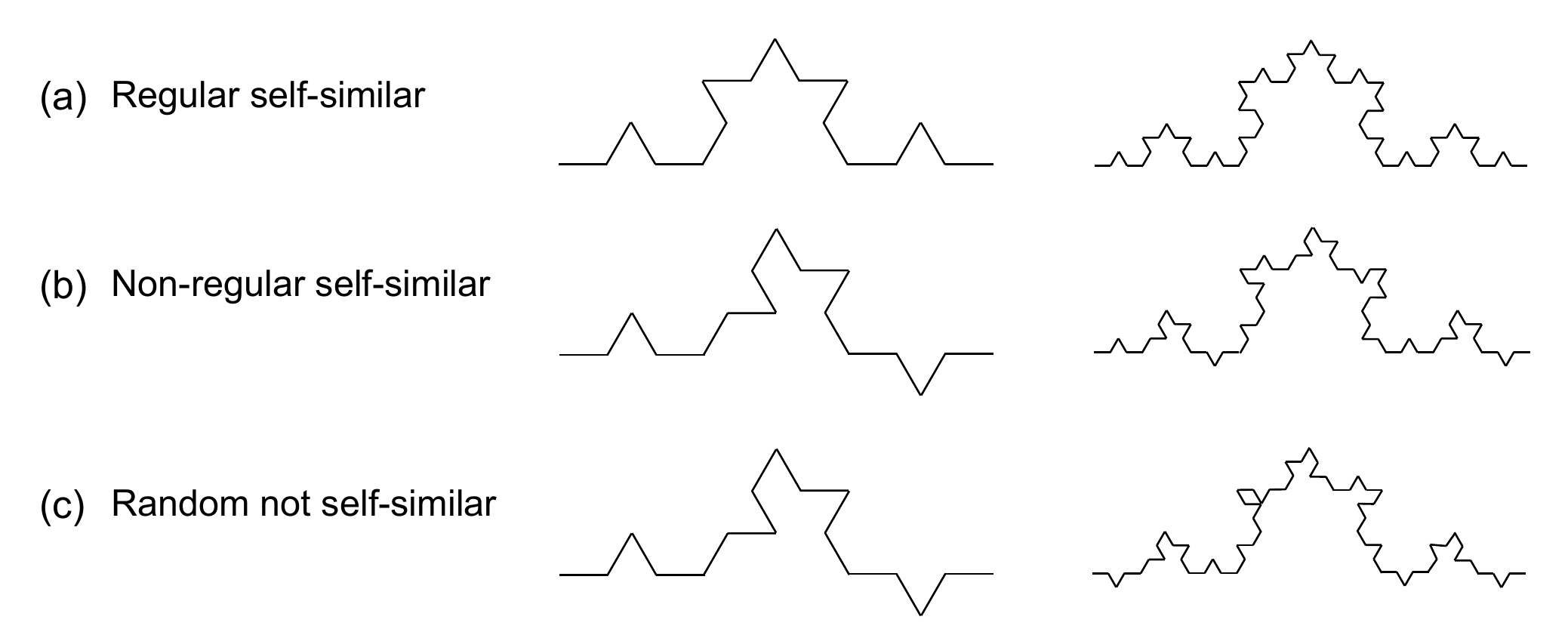}
		\caption{Different generation laws, in order to build a: (a) regular sequence, self-similar; (b) non-regular sequence self-similar and (c) random sequence, non self-similar.}
		\label{fig:rand}
	\end{figure}
	
	The dynamical responses to initial excitations induced by a moment and by a horizontal force for a set of curves generated with the triadic initiator using the random not self-similar pattern, Fig.\ref{fig:rand}(c), are shown in Fig.\ref{fig:Tkrand}. All terms in the sequence are composed by the assemblage of a well-defined geometry in decreasing scales but with different orientations in space. Fig.\ref{fig:rand}(a) represents the classical sequence of the Koch triadic. The Fig.\ref{fig:rand}(b) corresponds to a self-similar sequence of terms with a peculiar law of formation. It is similar to the Koch triadic without the symmetry displayed by the classical Koch triadic. The Fig.\ref{fig:rand}(c) is a non self-similar sequence, built with the triadic initiator properly reduced to fit the order of the term in the sequence but with a random orientation. 
	
	The dynamical approach proposed here leads to the following conclusions:
	
	\begin{enumerate}
		
		\item All curves belong to the Koch triadic family since the response to an initial displacement generated by the moment leads to a straight line as given by equation (2a) with angular coefficient $-0.1309$ corresponding to $F_M=D=1.2618$ which is the dimension of the Koch triadic Fig.\ref{fig:randM}.
		
		\item The response corresponding to an initial displacement induced by a horizontal force is a randomly distributed set of points for the case 2(c). A straight line obtained with the least square method applied to this set of points is a straight line with the same slope as in the previous case Fig.\ref{fig:randM}. Therefore, this sample consists of a random assemblage of a sequence of Koch triadic curves properly downscaled.
		
	\end{enumerate}

	\begin{figure}[h]
	\centering
	
	\subfigure[]{
		\includegraphics[width=0.45\textwidth]{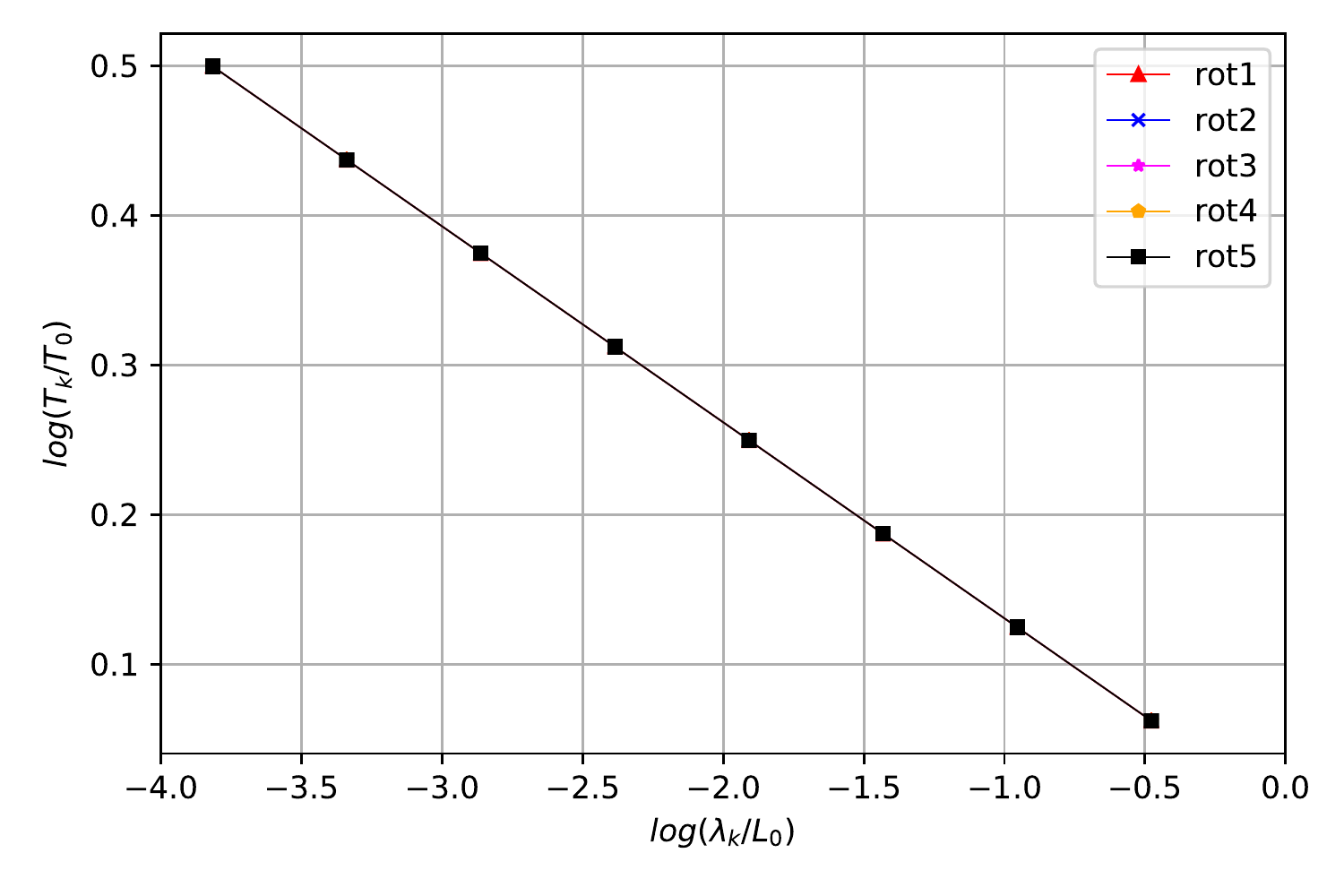}
		\label{fig:randM}
	}	
	\subfigure[]{
		\includegraphics[width=0.45\textwidth]{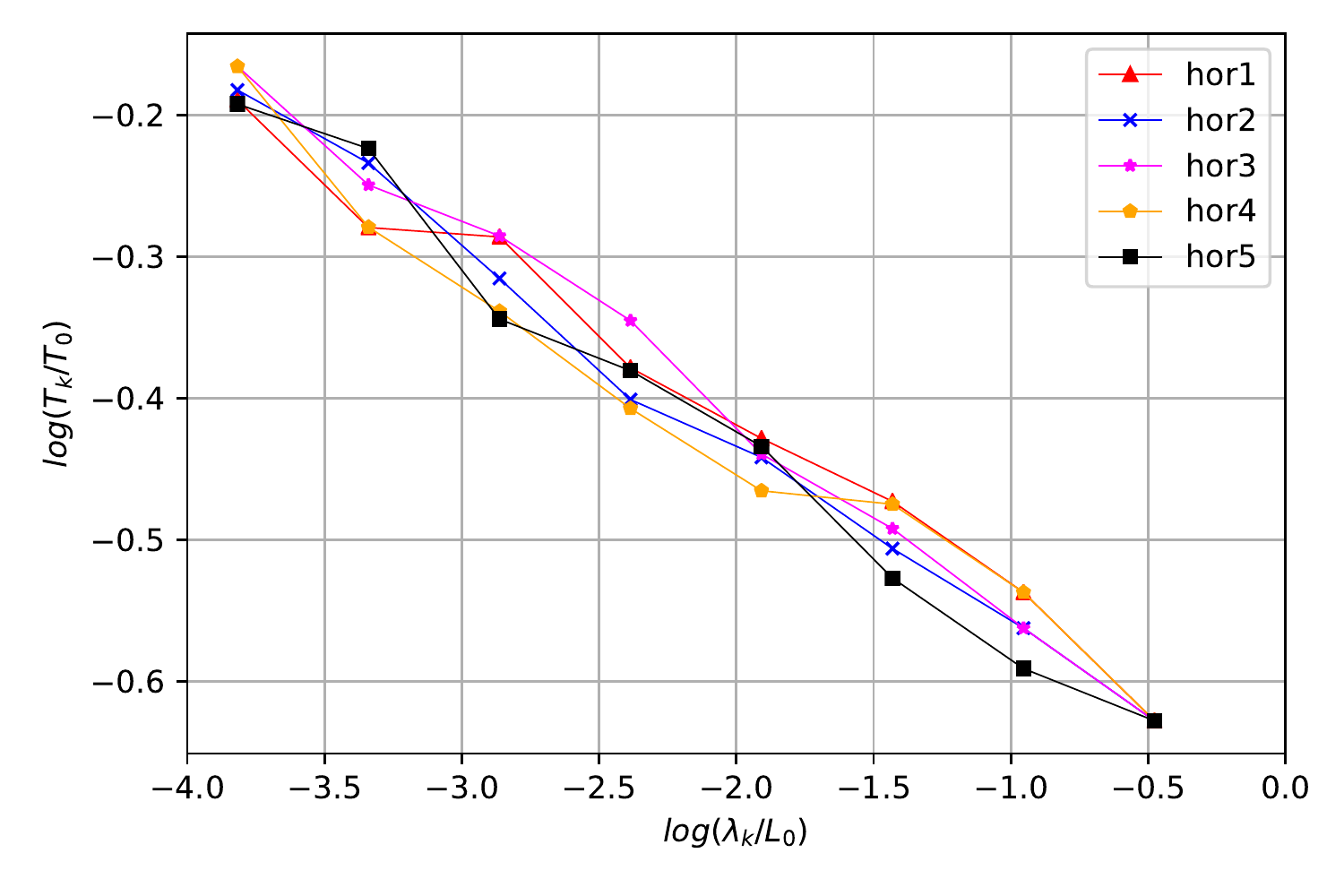}
		\label{fig:randH}
	}
	\caption{Log of the normalized period versus the log of the relative length of the reference element for five random triadic sequences. (a) Initial condition imposed by a moment at the free end. (b) Initial condition imposed by a horizontal force at the free end.}
	\label{fig:Tkrand}
	\end{figure}

	Classical methods to determine the fractal dimension are not able to find the fine structure of fractal curves. Additionally, the dynamical characterization of fractal curves provides a class of information not accessible to other methods, namely the material properties of the objects represented by the curves. More precisely, the fractal dynamic dimension can characterize a material object by combining its length with cross section geometry, specific mass and elastic properties. Particularly important is the consideration of the wire cross section geometry which plays an important role in the geometry of natural objects as trees and blood vessels. 
	
	To analyze the contribution of mass, material properties and cross section variations to the fractal dimension consider the equation \eqref{eq:Tk} and let the mass, the elastic modulus, and the moment of inertia of the cross section vary according to the following power laws:
	
	\[
	m_k/m_0=(\lambda_k/L_0)^\nu, \qquad 
	E_k/E_0=(\lambda_k/L_0)^\gamma, \qquad
	I_k/I_0=(\lambda_k/L_0)^\mu.
	\]
	
	Introducing the above relations in equation \eqref{eq:Tk} and after similar operations as before the following equation is obtained:
	
	\begin{equation}
		\log\left(\frac{T_k}{T_0}\right) = \frac{1-\overline{F}_{M,H,V}}{2} \log\frac{\lambda_k}{L_0}+\log \Omega_k \qquad \text{where} \quad \overline{F}_{M,H,V}=F_{M,H,V}+\mu+\gamma-\nu
		\label{eq:logTkMHV}
	\end{equation}

	\end{subequations}

\begin{figure}[h]
		\centering
		\includegraphics[width=0.5\textwidth]{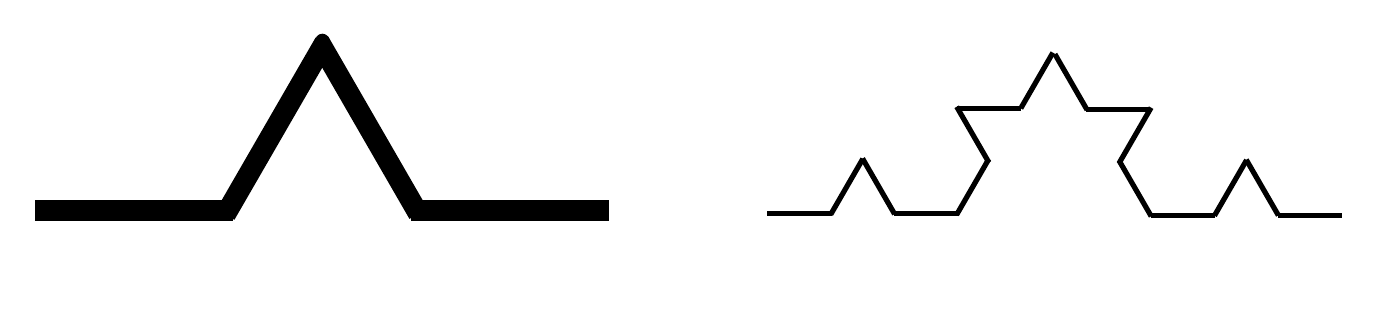}
		\caption{The dynamical fractal dimension may consider additional geometric properties of the elements in a sequence as for the case shown above $\overline{F}_{M,H,V}=(1+\mu-D)/2$.}
		\label{fig:varsec}
	\end{figure}

	Clearly the dynamical fractal dimension depends on a larger group of parameters as compared with the classical geometric approach. This is a particularly important result. Indeed, with an appropriate variation of the cross section the dynamical fractal dimension of a certain sequence may vanish despite of the fact that the classical geometric dimension is different from zero (Fig.\ref{fig:varsec}).

\section{The inverse problem}
	
	The basic ideas about the new approach to analyzing the complex fractal dimension of material objects were introduced in the previous section. However, a critical question for practical applications concerns the inverse problem. That is, given an element find its fractal dimension, if any. Since the dynamical dimension $\overline{F}_{M,H,V}$  depends on four parameters there is no unique solution. The classical fractal dimension in the sense of the Hausdorff metric is obtained if $\mu=0, \gamma=0$ and $\nu=0$. It is not the purpose of this paper to discuss the uniqueness problem posed by the addition of three new parameters to the classical fractal dimension. The main purpose here is to use the dynamic dimension to obtain the fractal dimension of a given sequence. 
	
	With the dynamical approach it is possible to determine the fractal dimension of a given sample by exploring variables that are not accessible to classical methods. In \cite{Bevi0,Bevi1,Bevi2,Bevi3} it was shown that the dynamical dimension of a sequence of subsamples obtained by successively cutting the previous one from the original sample led to its fractal dimension.
	
	In this section we present other ways to determine the fractal dimension of a given curve. 
	
	\subsection{The hidden sequence}
	
\begin{figure}[H]
		\centering
		\subfigure[]{
		\centering
		\includegraphics[width=0.4\textwidth]{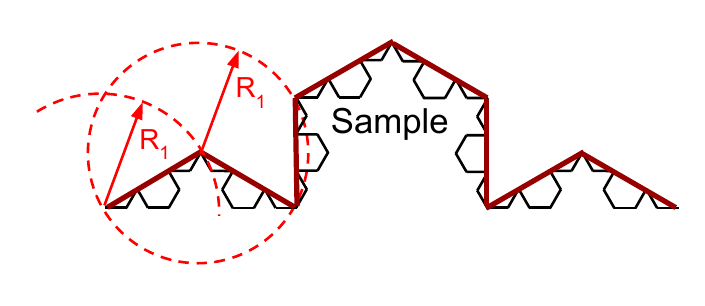}
			\label{fig:HS1}
		}	
		\subfigure[]{
		\includegraphics[width=0.4\textwidth]{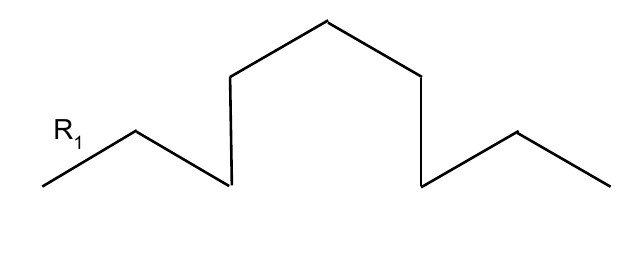}
		\label{fig:HS2}
		}	
		\subfigure[]{
		\includegraphics[width=0.4\textwidth]{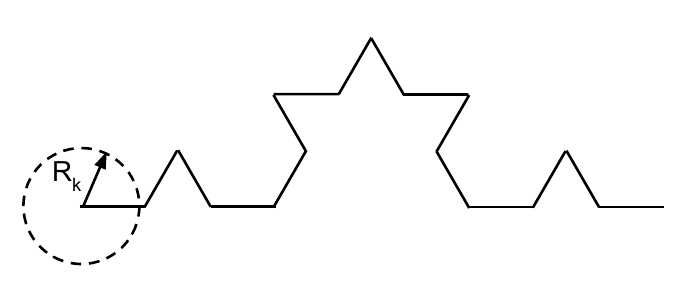}
		\label{fig:HS3}
		}		
		\subfigure[]{
			\includegraphics[width=0.4\textwidth]{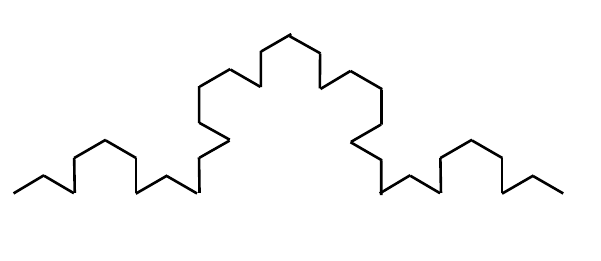}
			\label{fig:HS4}
		}	
		\subfigure[]{
			\includegraphics[width=0.4\textwidth]{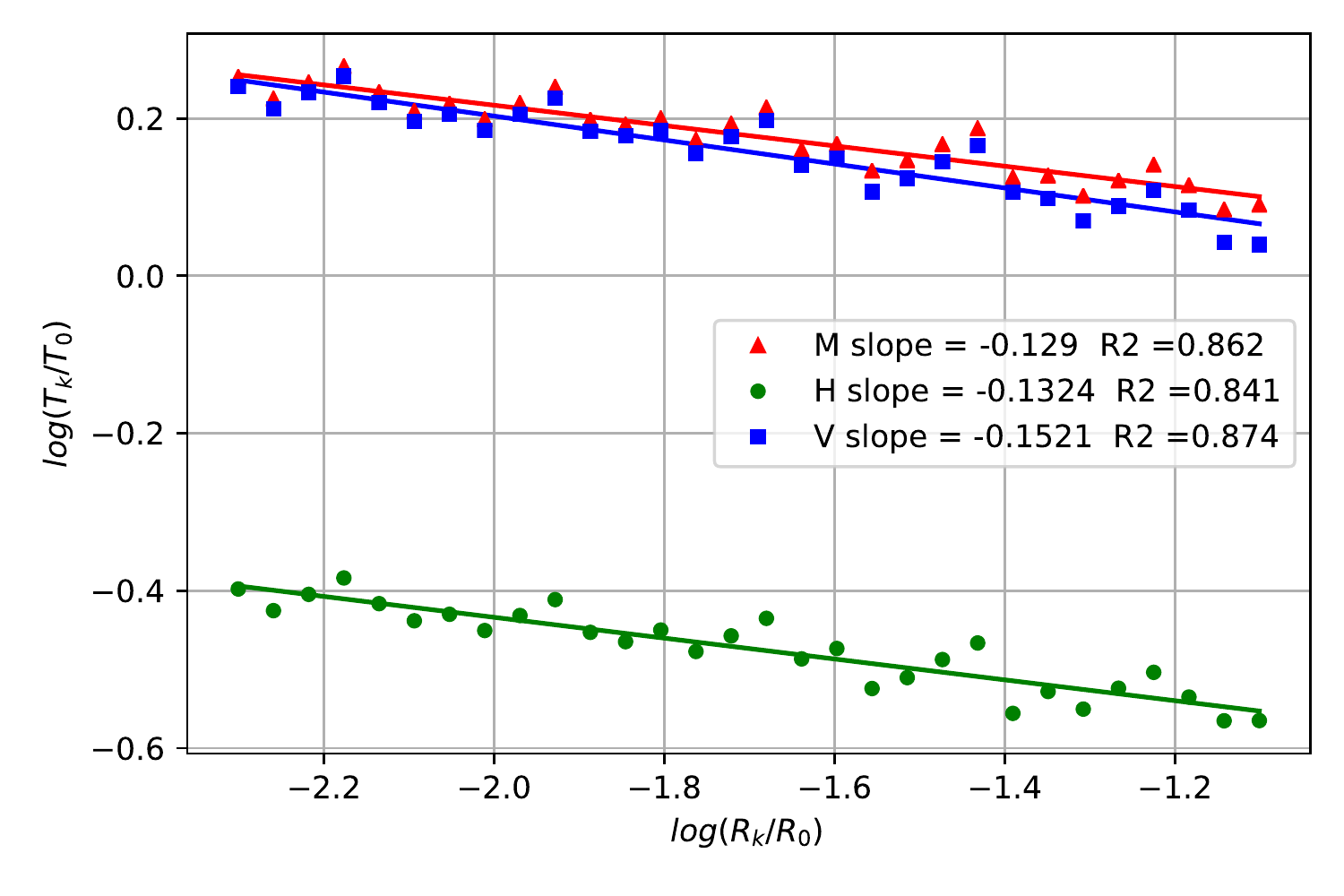}
			\label{fig:cterad}
		}
		\subfigure[]{
		\includegraphics[width=0.4\textwidth]{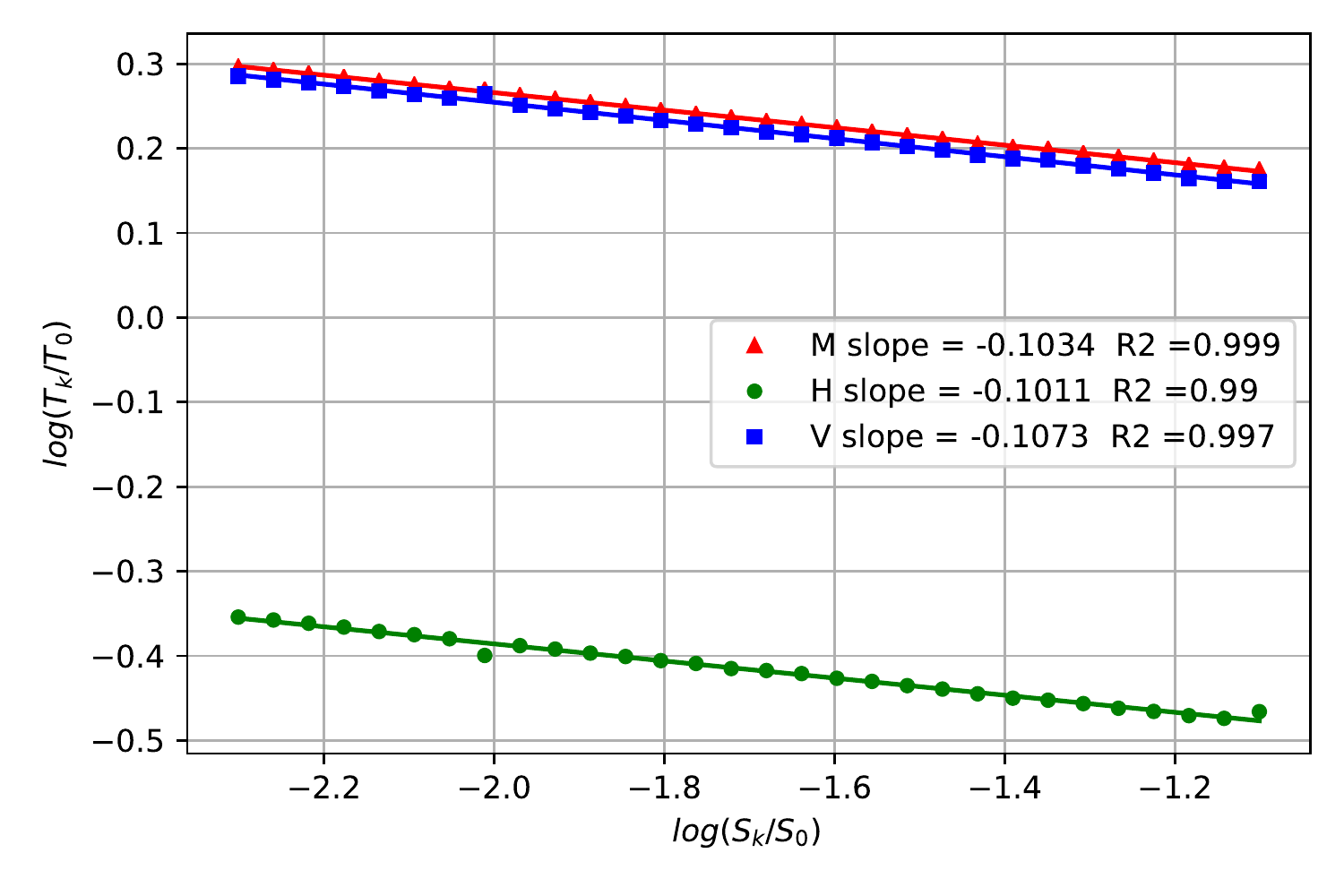}
		\label{fig:ctelen}
		}
		\subfigure[]{
		\includegraphics[width=0.2\textwidth]{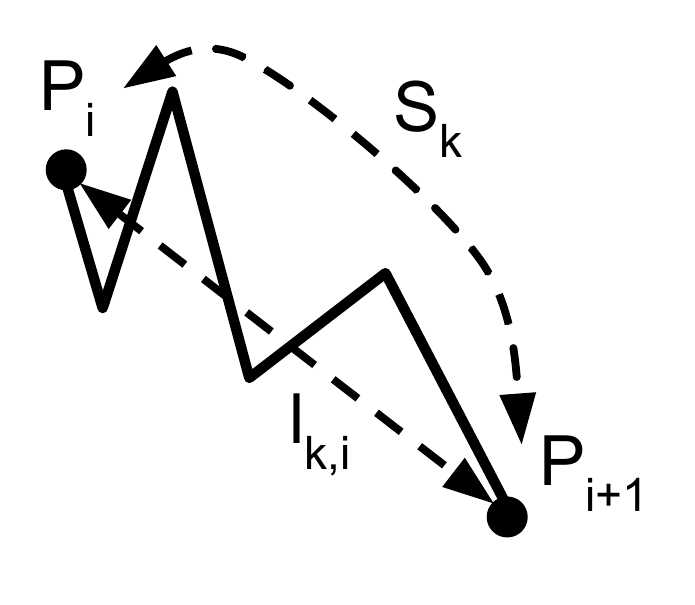}
		\label{fig:along}
		}	
		\caption{The sample (a) is decomposed in a sequence of curves each with a characteristic element length $R_k$ in (b,c,d). Dynamical behaviors associated with the constant radius approach (e), and with the constant length along the curve approach (f). Illustration of the constant along length approach to defining the element length $l_{k,i}$ (g)}
		\label{fig:Tkhidseq}
	\end{figure}

	Consider a plane curve $C$ as shown in Fig.\ref{fig:Tkhidseq}(a). There are two fundamental hidden sequences. The first consisting of a collection of curves $C_k (k=1,2,\dots,m)$ obtained by taking points $P_{k,1}, P_{k,2}, \dots P_{k,N_k}$ on $C$ such that the distance $ \overline{P_{k,i}P_{k,i+1}}= R_k$. For each value $R_k$ coresponds a term of the hidden sequence as shown in Fig.5(b,c,d). The second consists of a collection of curves $C_k$ defined by points on the curve $C$ such that the length $S_k$ measured on the curve $C$ is the same for any two consecutive points $P_{k,i}, P_{k,i+1}$, as shown in the Fig.\ref{fig:along}.
	
	The dynamical dimension is obtained from the hidden sequences. The results are shown in Fig.5(e) for the terms in a hidden sequence with constant segment length $R_k$ and in the Fig.5(f) for the terms in the hidden sequence with constant length along the curve $S_k$. The results are compared. The first option $R_k=constant$ for the given element of the hidden sequence displays a larger dispersion compared to the second criterion taking constant $S_k$. However, the solution obtained with $R_k=constant$ leads to a better approximation with $F_R\approx 1.26$ representing a relative error equal to 0.1\%.
	
	Now the expressions \eqref{eq:logTkM} and \eqref{eq:logTkHV} do not apply for the case $S_k=constant$ because for these groupings the elements $l_k$ are not constant. Let us call the elements of this type of grouping $l_{k,i}$. The expression \eqref{eq:Tk} must be rewritten:
	
	\[
	\frac{T_k^2}{T_0^2} = \sum_{i=1}^{N_k} \frac{l_{k,i}}{L_0}.
	\] 
	
	The $l_{k,i}$ elements are assumed to have all the same material properties and the same cross section dimensions. Now let $l_{k,i}=\overline{\alpha}_i l_k$:
	
	\[
	\frac{T_k^2}{T_0^2} = \sum_{i=1}^{N_k} \overline{\alpha}_i \frac{l_k}{L_0} = \left( \sum_{i=1}^{N_k} \frac{\overline{\alpha}_i}{N_k} \right) N_k \frac{l_k}{L_0}.
	\]
	
	With $\sum_{i=1}^{N_k} \overline{\alpha}_i/N_k=\alpha_k$ the expression above reads:
	
	\[
	\frac{T_k^2}{T_0^2} = \alpha_i N_k \frac{\lambda_k}{L_0},
	\]
	from which follows
	
	\[
	\log \left(\frac{T_k}{T_0}\right) = \frac{1}{2} (1-F^*) \log \frac{l_k}{L_0} \qquad \text{where} \quad F^* = \frac{\log(\alpha_k N_k)}{l_k/L_0}.
	\]
	
	It is also possible to write:
	
	\begin{equation}
		\log \left( \frac{T_k}{T_0} \right) = \frac{1}{2} [1-(F-\delta)] \log \frac{l_k}{L_0}
		\label{eq:Tkd}
	\end{equation}
	where
	\[
	\delta = \frac{\log(\alpha_k)}{\log(l_k/L_0)} \qquad \text{and} \quad F=\frac{\log(N_k)}{\log(l_k/L_0)}
	\]
	
	Since $\alpha_i \le 1$, $\delta \ge 0$ the angular coefficient of the line given in \eqref{eq:Tkd}, which is the dynamical dimension, is smaller than the corresponding classical fractal dimension $D$. The approach taking the approximation sequence constructed with points $P_i$, $P_{i+1}$ on the sample leads to a shorter total length since $\overline{P_i P_{i+1}} \le S_i$ for all $i$, where $S_i$ is the length measured on the curve. Therefore, $\alpha_k$ is less than one for all cases and $\delta$ is always positive. This means that $F^*$ is always less than the fractal dimension $F$. For the present case we find $F^*=1.20$ leading to the value of $\delta =0.03$.
	

	The function $\log(T_k/T_0)=f\left(\log(\lambda_k/L_0)\right)$ obtained from the virtual sequence with the first option $R_k=constant$, $k=1,2,\dots,n$ presents a greater dispersion as compared with the second criterion taking $S_k=constant$. However, the solution obtained taking $R_k=constant$ leads to a good approximation to the Hausdorff dimension, namely $D=F\approx1.26$ with a relative error equal to 0.1\%. The approximation with $S_k=constant$ leads to $F^*\approx1.20$. But $F^*$ does not represent the Hausdorff fractal dimension, it is a peculiar fractal dimension associated with the dynamical behavior of such particular oscillators sequence. 
	
	\subsection{Variation of the physical properties of the hidden sequence}
	
	Instead of determining the fractal dimension of a given sample it is often convenient to find the upper and lower bounds of the hidden dimension.
	
	Now the periods of the oscillators corresponding to the hidden sequence depend on the moment of inertia of the wire cross-section. If the inertia of the cross-section varies according to the law $I_k=I_0(\lambda_k/L_0)^\mu$ where $\lambda_k$ is the length of the characteristic element, the period can be determined by the equation:
	
	\begin{equation}
	\log\frac{T_k}{T_0} = \frac{(1-\mu-D)}{2} \log\left(\frac{\lambda_k}{L_0}\right)+\log\Omega_k
	\label{eq:Tkmu}
	\end{equation}   
	
\begin{figure}[h]
		\centering
		
		\subfigure[]{
			\includegraphics[width=0.45\textwidth]{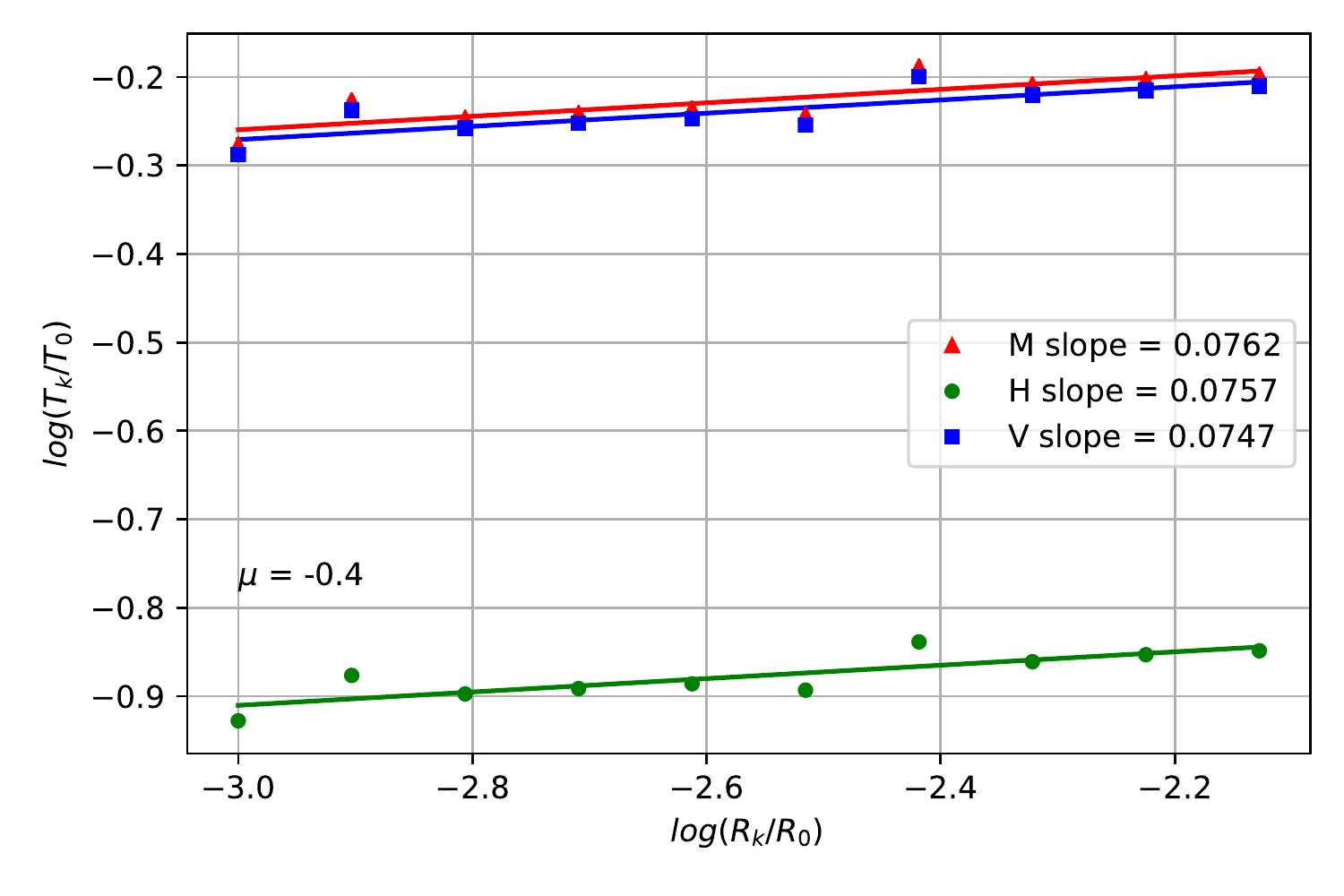}
			\label{fig:mu04}
		}	
		\subfigure[]{
			\includegraphics[width=0.45\textwidth]{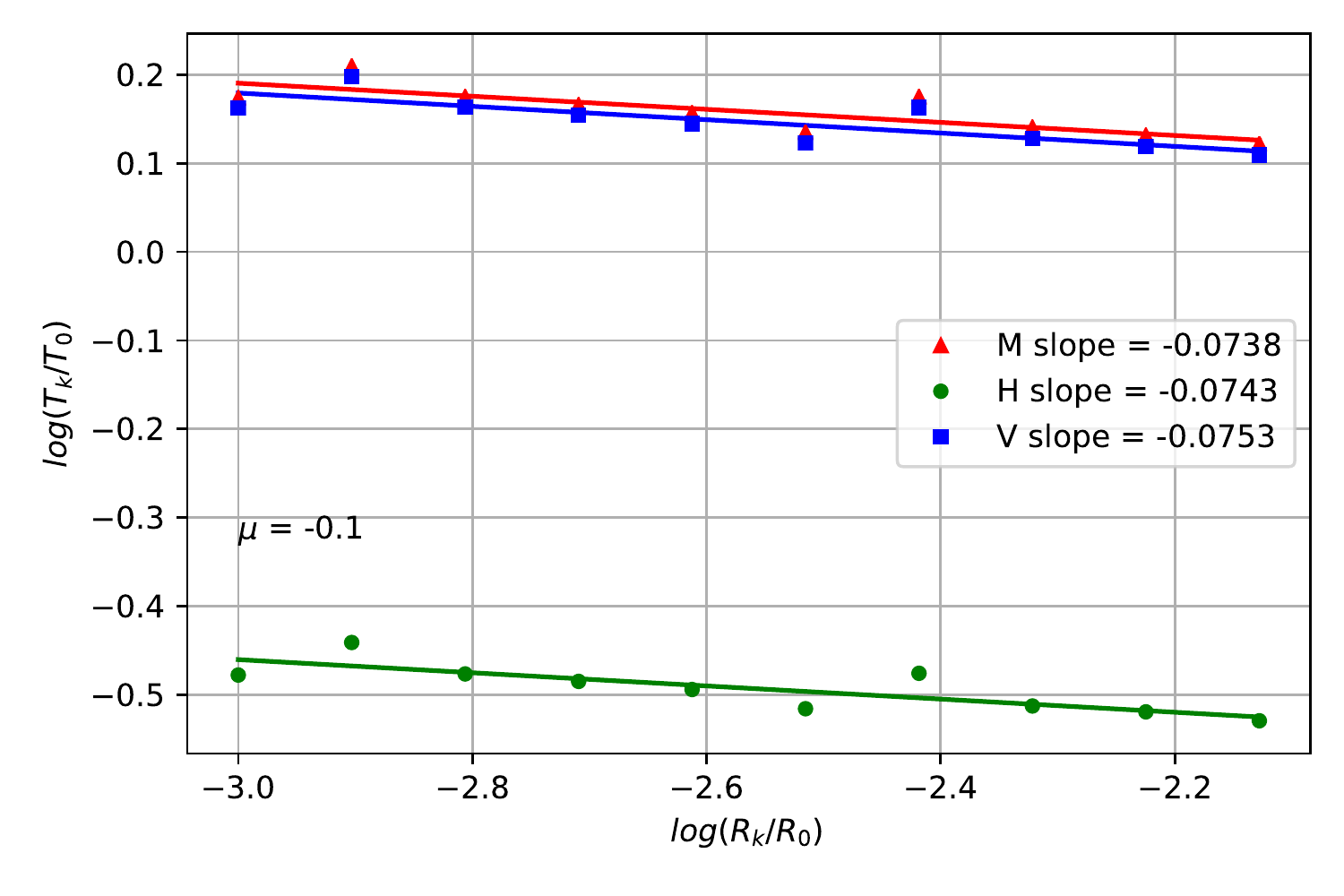}
			\label{fig:mu01}
		}
		\caption{Period as function of the characteristic element $\lambda_k$ for two sequences consisting of oscillators with different natural frequencies.}
		\label{fig:mu}
	\end{figure}

	This corresponds to an approximated straight line with angular coefficient $(1-\mu-D)/2$. The fractal dimension $D$ is a fixed value depending on the geometry of the plane curve. Let us determine the evolution of $\log(T_k/T_0)$ for different values of $\mu$ as shown in Fig.\ref{fig:mu}. For $\mu=-0.4$ the function \eqref{eq:Tkmu} has a positive slope and for $\mu=0.1$ it has a negative slope. 
	
	Now if for a given value $\mu=\mu^*$ the slope is equal to zero, $(1-\mu^*-D)=0$, the fractal dimension is determined, $D=1-\mu^*$. From the experiments above it is possible to say that $1.4>D>1.1$. Therefore, it is possible to determine the upper and lower bounds of the fractal dimension. The above results provide a first approximation of the range where the fractal dimension should be found. More accurate results can be obtained by adjusting the values of $\mu$ and observing the change in sign of the angular coefficient. The specific value of the exponent $\mu^*$ defining the characteristic of the wire cross-section for each sequence defines the characteristic transition between dynamical responses of the sequence of oscillators. That is, if the exponent that governs the geometry of the cross-sections keeps the periods constant for the sequence of oscillators, the corresponding fractal dimension is obtained from $\mu^*$, provided that the other properties are kept constant. 
	
	\subsection{Variation of the mass of the oscillators}
	
	Instead of defining a hidden sequence it is possible to find the fractal dimension of a plane curve through the dynamical response of the given sample by modifying the characteristics of the oscillator. The fractal dynamical dimension depends on the mass attached at the free end of the harmonic oscillator as shown in equation \eqref{eq:logTkMHV}. Therefore, the inverse problem can be easily solved by analyzing the responses of a sequence where all terms preserve the same geometry, but the mass varies according to an appropriate law.
	
\begin{figure}[h]
	\centering
	
	\subfigure[]{
		\includegraphics[width=0.45\textwidth]{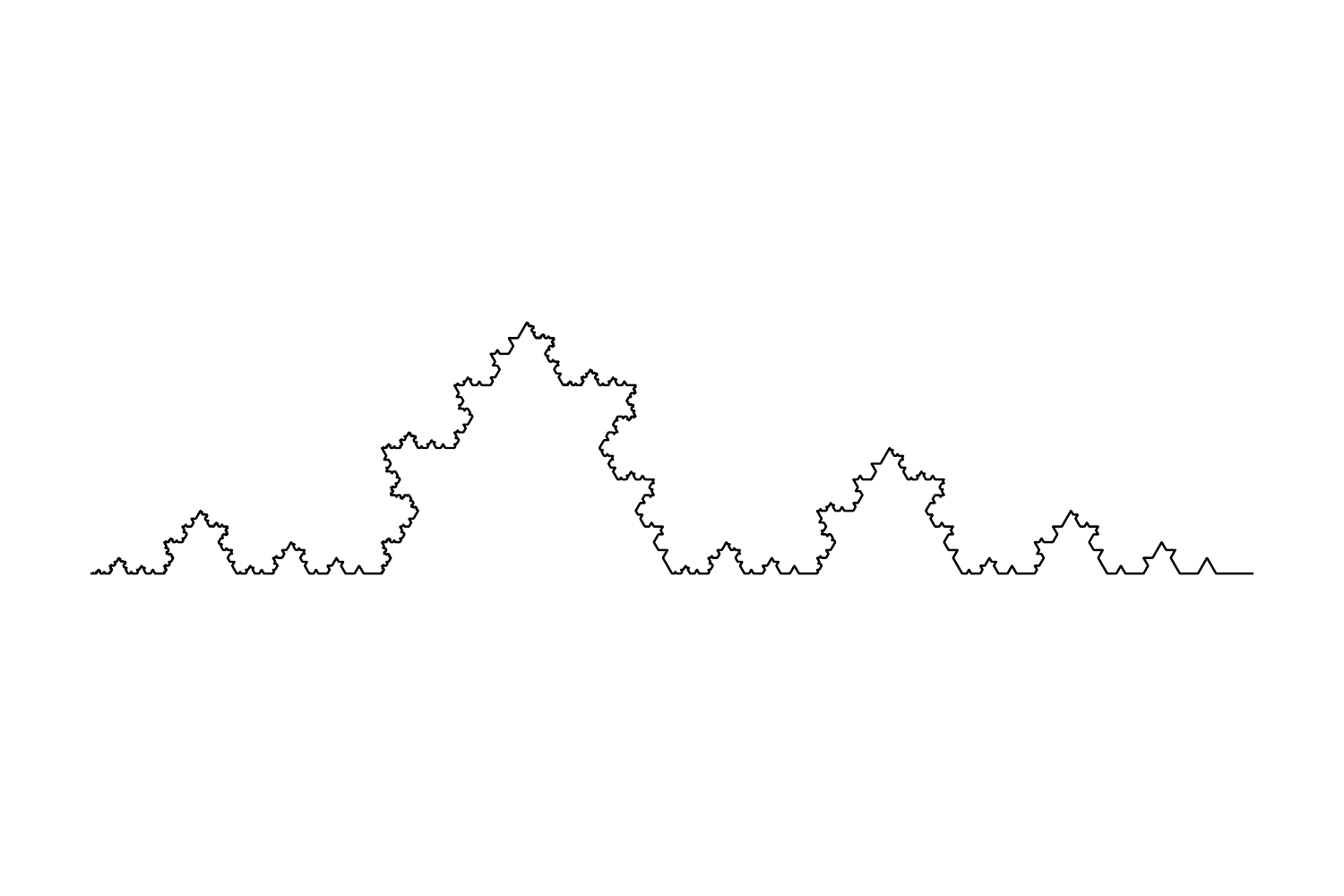}
		\label{fig:kochmod}
	}	
	\subfigure[]{
		\includegraphics[width=0.45\textwidth]{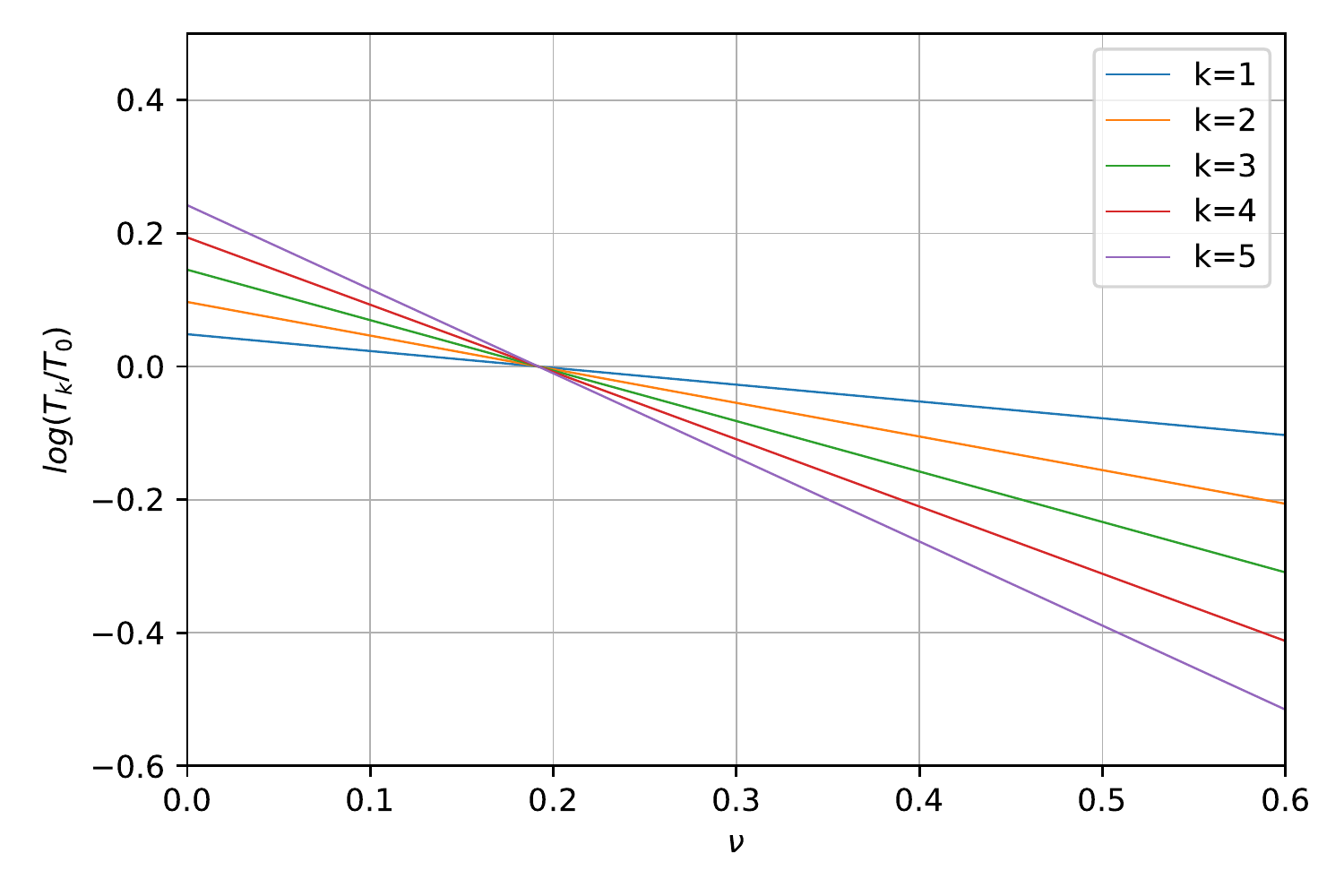}
		\label{fig:mass}
	}
	\caption{(a) Determination of the geometric fractal dimension of a curve assembled with the initiator of the Koch triadic at different scales. (b) Variation of the relative period with the mass coefficient $\nu$. Each line corresponds to a given order of the geometry, namely $k=1,2,3,4,5$.}
	\label{fig:masstot}
	\end{figure}

	Consider the curve in Fig.\ref{fig:kochmod}. It is a curve assembled with the Koch initiator on a given scale with the characteristic segment given by $\lambda_0$. Now for each reference mass $m_i$ ($i=1,2,\dots,n$) let us build a sequence of masses $m_{i,\nu}$ for each $\nu=\nu_1,\nu_2,\dots,\nu_m$  according to the power law:
	
	\begin{equation}
		\frac{m_{k,\nu}}{m_0} = \left( \frac{\lambda_k}{L_0} \right)^\nu, 	\quad \nu \ge 0.
		\label{eq:mnu}
	\end{equation}
	
	For the initial displacement imposed by a moment the relative period is given by:
	
	\begin{equation}
		\log \frac{T_k}{T_0} = \frac{(1+\nu-D)}{2} \log\left(\frac{\lambda_0}{L_0}\right)
		\label{eq:Tknu}
	\end{equation}
	
	The responses of each series of oscillators for five characteristic values of $m_k$ are given as a function of $\nu$ in Fig.\ref{fig:mass}. They all cross the line $\log(T_k/T_0)=0$ at $\nu=0.1918$. Since for this point $(1+\nu-D)=0$ according to \eqref{eq:Tknu} the fractal dimension of the curve is determined, namely $D=1.1918$.
	
	This method is very accurate and uses the sample itself to determine its fractal dimension. Besides, the solution exposed here suggests that the fractal dimension can be determined experimentally. It is not difficult to build a harmonic oscillator with a spring bent according to the geometry determined by the fractal curve. The dynamical responses of the oscillator for a set of masses $m_\nu, \nu=\nu_1,\nu_2,\dots,\nu_m$ varying according to a power law as given in \eqref{eq:mnu} can be used to determine the fractal dimension with equation \eqref{eq:Tknu}.
	
\section{Conclusion}

	In the previous sections, the application of a new method to determine the fractal characteristics of fractal curves was shown. The new method uses the dynamical behavior of simple oscillators to determine the particular characteristics of fractal curves. The elastic element of the oscillator is a linear spring simulating the geometry of the fractal curve. As the linear oscillator on a plane has three degrees of freedom the information obtained from the dynamical behavior is equally more complete than the information obtained with the geometric approach. This fact is clear from the results obtained for the case of random fractals as explained above.
	
	The inverse problem, that is, the identification of possible fractal characteristics in a given curve can be explored with the dynamical method with different approaches since three material characteristics are involved in the process. The geometry of the wire cross-section together with its respective specific mass and elastic properties play a decisive role in determining the fundamental frequencies of the oscillators. Therefore, the fractal dynamic dimension is more suitable for characterizing material objects than the classical approach.
	
	To solve the problem posed by the influence of the cross-section on the dimension of real objects it is suggested to analyze the cross-section variation as an independent variable. This, however, is not satisfactory, since the coupling of the two independent dimensions is incomplete. The dynamical approach automatically maintains coupling providing consistent characterization of material objects. Networks of blood vessels and branch bifurcations in plants are characteristic examples of these material objects \cite{Barros}.
	
	A considerable advantage of the dynamical approach consists of determining the fractal dimension from a single sample in a fractal sequence, provided that the given sample contains sufficient information. With the sample it is possible to build a sequence of curves, that we call the hidden sequence, with an appropriate criterion as shown in the section 2.1. 
	
	The solution of the identification problem can also be obtained by modifying other variables that interfere with the dynamical response, such as mass. This approach can lead to the solution with the help of the given sample coupled to an appropriated sequence of masses composing a corresponding sequence of oscillators. With this method the original sample remains undisturbed, as shown in section 2.3. 

	Therefore, it is also possible to obtain the fractal dimension of a given sample through experimental procedures with the help of a linear oscillator. The oscillator is assembled with an elastic spring in the shape of the sample. The sequence of the fundamental frequencies, the first harmonic, corresponding to an appropriated sequence of masses discloses the fractal dimension.

	The method presented here can be easily extended to the case of curves in three-dimensional space. The dynamical approach to analyzing fractal curves is just beginning. It is rich enough to open up a consistent theoretical and practical investigation of the particular geometric formation of curves and surfaces. The dynamical dimension is richer than the geometric fractal dimension and should be seen as an independent theory. Real objects can be characterized more consistently through the dynamical dimension as it incorporates in the fractal dimension not only the curve geometry  but also complementary material characteristics such as density, elastic properties and the geometry of the cross section.

\section{Acknowledgements}

We acknowledge the CNPq (National Research Council) and the ANP (Brazilian National Agency for Petroleum, Natural Gas and Biofuels) for the support given to the authors through the Senior Research Grant and the contract ANP 5850.0104954.17.9: "Theoretical and experimental research on the application of physical methods for the mitigation of mineral incrustations", respectively.


\end{document}